# Novel miniaturized low-pass filters from artificial transmission line structures

Yu He, Xin Wu and Chang-jun Liu*

*School of Electronics and Information Engineering, Sichuan University, Chengdu 610064, China*

A novel design on miniaturized low-pass filters from artificial transmission line structures is proposed. Reactance components, i.e. capacitors and inductors, are directly constructed from microstrip structures based on electromagnetic simulation and parameter extraction. The constructed reactance components are integrated to build low-pass filters. The results of a fifth-order Butterworth, 0.25, 0.5, and 1 dB ripple Chebyshev filters, which are realized by the proposed method, are presented. The proposed low-pass filters are compact and their sizes are about 20% of conventional microstrip low-pass filters. The simulation agrees well with the measurements, which shows the validation of the proposed design method. The miniaturized low-pass filters may be applied to many microwave systems.

**Keywords:** microstrip filter; miniaturization; low-pass filter; artificial transmission line; reactance components; parameter extraction

## 1. Introduction

Microstrip low-pass filters have been widely used in microwave system due to the low fabrication cost.[1] Miniaturization of microstrip filters has received substantial attention and many works are devoted to reducing the filter size.[2,3] Artificial transmission line reported in the literature is used to reduce the size of microwave components, such as power dividers, couplers, and baluns.[4–6] In this letter, by utilizing the artificial transmission line, a novel miniaturized low-pass filter is designed based on parameter extraction from the constructed reactance components. Meandered lines and parallel-plated capacitors are used to replace lumped inductors and capacitors, respectively.

Filter designs at microwave frequencies are difficult to realize with lumped components because the wavelength becomes comparable with the physical dimensions of filter structure, resulting in more losses. Thus, distributed transmission line elements are of importance for designing practical microwave filters. The conventional approach to converting lumped elements into distributed elements is relying on Richard's transformation, while Kuroda's identities can be applied to separate filter elements using transmission line sections. However, they highly increase the filter size, though such transmission line sections do not affect the filter response. The large filter size makes it inapplicable in current wireless communication which demands miniaturization and integration.

A new methodology of filter design is proposed to tackle the aforementioned problem. Reactance components, i.e. capacitors and inductors, can be directly constructed from microstrip structures based on electromagnetic simulation and parameter extraction. The constructed reactance components are integrated to form a desired filter. Utilizing this method to design a filter applies well to a low-band microwave. The circuit size is more compact compared to conventional distributed element filters.

---

*Corresponding author. Email: cjliu@ieee.org

## 2. Design concept and methodology

According to the aforementioned filter design concept, the design steps of a low-pass filter are as follows: (a) select the normalized filter parameters to meet the design criteria; (b) obtain the filter parameters $L_i$ and $C_i$ with frequency transformation and impedance transformation; (c) utilize electromagnetic simulation to construct the equivalent microstrip structure and extract the equivalent element parameters; and (d) integrate and optimize the microstrip structure.

There are two possible realizations for the generic normalized low-pass filter. Here, we choose the first element of the low-pass filter to be an inductor connected in series with the source. The equivalent circuit of a low-pass filter is given in Figure 1.

In order to illustrate the design procedure for this type of filter, a design of a five-pole low-pass filter with a Butterworth response and a cutoff frequency at 1.8 GHz is described as follows. The element values of Butterworth responses are $g_1 = g_5 = 0.618$, $g_2 = g_4 = 1.618$, and $g_3 = 2$. With frequency transformation and impedance transformation, we can get $L_1 = L_5 = 2.73$ nH, $C_2 = C_4 = 2.86$ pF, and $L_3 = 8.84$ nH. Meandered lines and parallel-plated capacitors are constructed to equate with these values. The full-wave electromagnetic simulator IE3D [7] is used to extract the element values from the equivalent circuit model. The constructed inductor based on a microstrip meandered line and its corresponding equivalent-lumped circuit model are shown in Figure 2(a) and (b), respectively. Meandering of transmission lines could be the most straightforward way for size reduction.[8–10] Referring to the equivalent model in Figure 2(b), the inductor $L$ represents meandered line inductor, while $C_1$ and $C_2$ represent parasitic capacitance. For the sake of simplicity, $C_1$ and $C_2$ are assumed to be identical. The Y-matrix of the equivalent model is written in the form

$$[Y] = \begin{pmatrix} Y_{11} & Y_{12} \\ Y_{21} & Y_{22} \end{pmatrix} = \begin{pmatrix} j\omega C_1 + \dfrac{1}{j\omega L} & -\dfrac{1}{j\omega L} \\ -\dfrac{1}{j\omega L} & j\omega C_2 + \dfrac{1}{j\omega L} \end{pmatrix} \quad (1)$$

$L$ and $C$ can be obtained by the extraction of the imaginary part of Y-matrix

$$L = \frac{1}{2\pi f \operatorname{Im}(Y_{21})} \quad (2)$$

$$C_1 = C_2 = \frac{\operatorname{Im}(Y_{11}) + \operatorname{Im}(Y_{21})}{2\pi f} \quad (3)$$

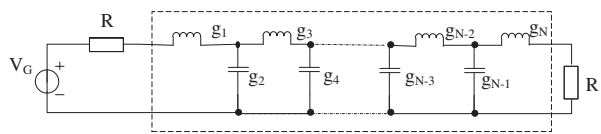

Figure 1. Equivalent circuit model of a low-pass filter.

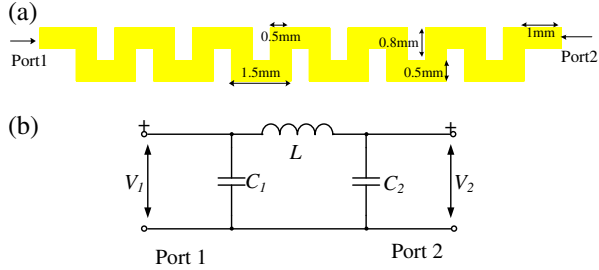

Figure 2. Constructed inductor based on a microstrip meandered line. (a) Circuit schematic. (b) Equivalent-lumped circuit model.

According to (2) and (3), *L* and *C* are the functions of frequency. Therefore, the equivalent parameters *L* and *C* can only be obtained with a specific frequency substituted into (2) and (3). The *Y*-matrix of meandered lines is extracted with IE3D simulation from 0 to 5 GHz. The length and width of the meandered line are optimized based on the design parameter. At the cut-off frequency 1.8 GHz, $L = 7.20$ nH and $C = 0.40$ pF. Its equivalent inductance is $L_{\text{eff}} = 8.84$ nH, which corresponds with the design value.

The method can also be applied to construct a parallel-plated capacitor. The constructed capacitor from microstrip structure and its corresponding equivalent-lumped circuit model are shown in Figure 3(a) and (b), respectively. The *Y*-matrix of the equivalent model can be given by

$$[Y] = \begin{pmatrix} Y_{11} & Y_{12} \\ Y_{21} & Y_{22} \end{pmatrix} = \frac{1}{j\omega L + \frac{1}{j\omega C}} \begin{pmatrix} 1 & 1 \\ 1 & 1 \end{pmatrix} \quad (4)$$

*L* and *C* can be obtained by the extraction of the imaginary part of *Y*-matrix

$$L = \frac{\text{Im}\left(\frac{1}{Y}\right) + f \frac{\partial \text{Im}\left(\frac{1}{Y}\right)}{\partial f}}{4\pi f} \quad (5)$$

$$C = \frac{-1}{\pi f \left[\text{Im}\left(\frac{1}{Y}\right) - f \frac{\partial \text{Im}\left(\frac{1}{Y}\right)}{\partial f}\right]} \quad (6)$$

Similarly, with parameter extraction by simulation at cutoff frequency 1.8 GHz, we get $C = 1.48$ pF and $L = 2.57$ nH.

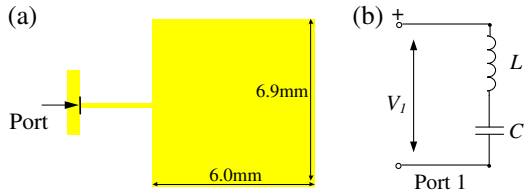

Figure 3. Constructed capacitor from microstrip structure. (a) Circuit schematic. (b) Equivalent-lumped circuit model.

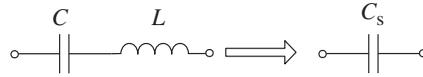

Figure 4.  Equivalence of series *LC* resonant circuit.

The series *LC* resonant circuit can be equivalent to a series capacitor, as shown in Figure 4.

The impedance of series *LC* resonant circuit is

$$Z_{LC} = j\omega L + \frac{1}{j\omega C} \tag{7}$$

The impedance of single series capacitor is

$$Z_C = \frac{1}{j\omega C_S} \tag{8}$$

Let $Z_{LC} = Z_C$, we can obtain the value of $C_S$ with $C = 1.48$ pF and $L = 2.57$ nH substituted into the formula (9). We get $C_S = 2.88$ pF which is quite close to the design value 2.86 pF.

$$C_S = \frac{C}{1 - \omega^2 LC} = \frac{C}{1 - 4\pi^2 f^2 LC} \tag{9}$$

The constructed meandered lines and parallel-plated capacitors are used to replace the original lumped elements as shown in Figure 5.

## 3. Experiments

Following the design procedure, the filters developed in this study are fabricated on F4B-2 substrate with a thickness of 1 mm, a dielectric constant $\varepsilon_r$ of 2.65, and a dielectric loss of 0.004. The sizes of the four filters are 38.0 mm × 18.6 mm, 46.2 mm × 18.3 mm, 50.7 mm × 18.3 mm, and 58.4 mm × 18.5 mm, respectively. The Agilent N5230A network analyzer is used to measure $|S_{11}|$ and $|S_{21}|$ from 0 to 10 GHz. The simulated and measured results are shown in Figure 6.

As shown in Figure 5(a), the maximum insertion loss of the five-pole Butterworth filter is as low as 0.5 dB at its operation frequency and the rejection band is from 1.8 to 4.5 GHz. The pass-band ripples of the three Chebyshev filters in Figure 5(b)–(d) are lower than 0.21, 0.29, and 0.47 dB, respectively. The simulated S-parameters and the measured results of the proposed low-pass filters are in good agreement.

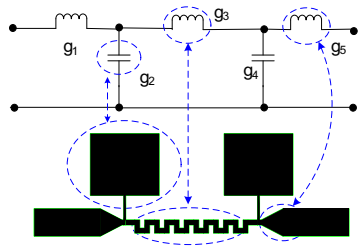

Figure 5.  Reactance components replace the lumped elements.

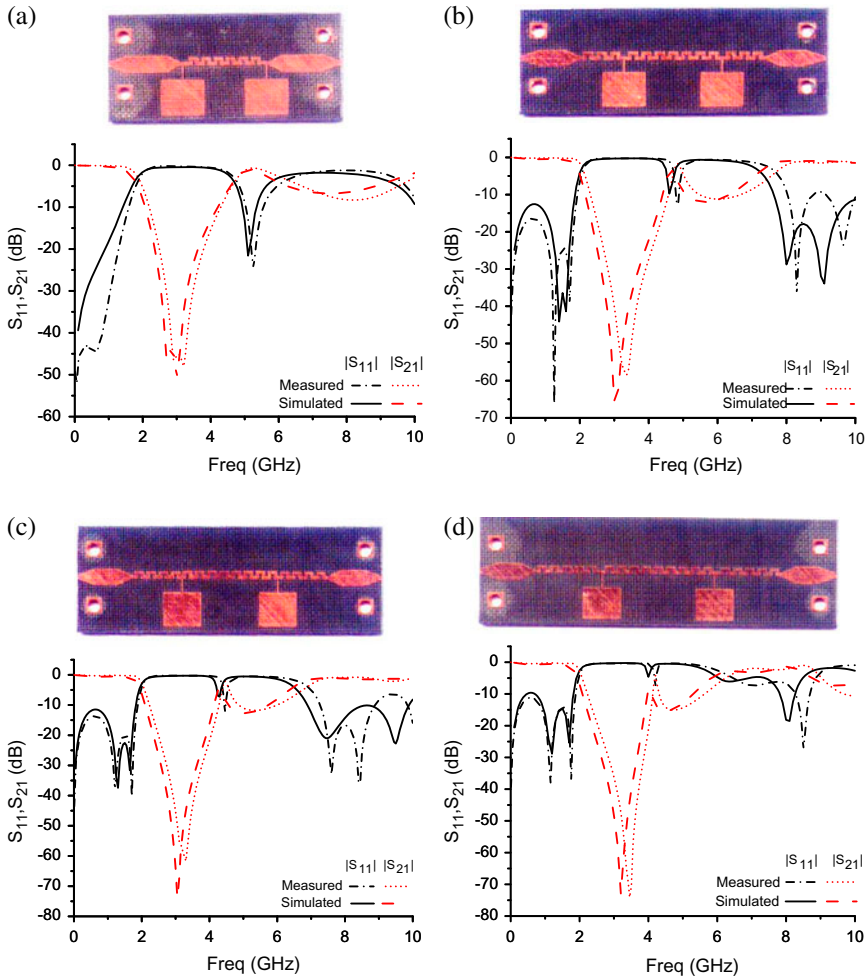

Figure 6. Low-pass filters with simulated and measured results. (a) The fifth order Butterworth filter. (b) The fifth order 0.25dB ripple Chebyshev filter. (c) The fifth order 0.5dB ripple Chebyshev filter. (d) The fifth order 1dB ripple Chebyshev filter.

## 4. Conclusion

In this paper, a novel miniaturized low-pass filter design methodology based on constructing reactance components by artificial transmission line is proposed. Reactance components are directly constructed based on electromagnetic simulation and parameter extraction. The lumped elements of the filters are then replaced by meandered lines and parallel-plated capacitors. The proposed low-pass filter is compact and its size is about 20% of a conventional filter. Four filters with different characteristics are designed for the validation check, and the measured results agree well with the simulations.

## Funding

This work was supported in part by the 973 program 2013CB328902, NFSC 61271074, and NCET-12-0383.